# ON ABNORMAL ABSORPTION OF HADRON COMPONENT OF EAS CORES IN LEAD AND POSSIBLE EXPLANATIONS

L.G. Sveshnikova[a], A.P. Chubenko[b], R.A. Mukhamedshin[c], N.S. Popova[a],
N.M. Nikolskaya[b], V.I. Yakovlev[b].

a) Skobeltsyn Institute of Moscow State University, Leninskie gory, Moscow, 119992 Russia
b) Lebedev Physical Institute, RAS, Leninskii prospekt, 53, 117924 Russia
c) Institute for Nuclear Research, RAS, prospekt 60-letiya Oktyabrya 7A, Moscow, 117312 Russia

We confirm the result obtained many years ago at Tien-Shan mountain station with the large 36-m$^2$ lead calorimeter that in extensive air showers (EAS) with energies of few PeV the attenuation of core energy deposit in lead becomes slower than it could be predicted by modern codes. It is shown that this effect is connected with the appearance of the excess of abnormal cores with a large ionisation released in lower layers of the calorimeter and these abnormal EAS cores are most probably produced by high-energy muon groups. A few hypotheses of the excess of muon reach EAS cores are considered. To study the absorption of EAS hadrons and muons in a lead ionization calorimeter, the EAS development in the atmosphere was simulated in the framework of the CORSIKA+QGSJET code whereupon the passage of hadrons and muons through a lead calorimeter has been modeled with using the FLUKA transport code.

## Introduction

The development of the so-called extensive air showers (nuclear-electron-photon cascades in the atmosphere), or EAS, originated by high-energy primary cosmic rays have been studied during half a century. However, only a few EAS arrays include the hadronic detectors to study the EAS hadron component. This component degrades on the path to the sea level very strongly, so only a tiny (1% – 4%) part of the EAS primary energy depending on the observation level and incident angle is carried by hadrons; the rest of the energy is transformed to the electromagnetic and muon components. However, just the hadron component determines the rate of EAS development in the atmosphere; besides, it is very sensitive to such important parameters of interaction as cross sections and inelasticity coefficients of hadrons. The oldest array where the EAS hadron component was studied in detail at mountain levels is the EAS array with a big lead ionization calorimeter (BIC) at the Tien Shan mountain station of the Lebedev Institute of Physics (Moscow) sited at a height of 3340 m a.s.l. Data obtained by this array continue to be unique due to a calorimeter's large area of 36 m$^2$ and lead large thickness of 850 g/cm$^2$ (i.e., five proton's mean free paths).

One of unexplained results obtained with the help of the BIC is the so-called long-flying component [1]. First data concerning the registration by the ionization calorimeter of long-flying component in EAS cores have been published in 1974 [2]. It was found that the EAS-core energy absorption rate in lead diminishes steadily with increasing the registered hadron energy. In 1975, E.L.Feinberg has assumed [3] that charmed particles produced in the points of interaction of hadrons with lead nuclei in the calorimeter could be responsible for the delay of the energy flux absorption in EAS cores. After charmed particles parameters had been specified, I.Dremin at al. [4,5] showed that the experimental data on the hadronic component in EAS cores, which had been obtained at energies higher than 30 TeV, could be explained under the assumption that the charmed-particle generation cross section attains ~30% of the inelastic proton-nucleus cross section. In addition, these particles could carry away a significant fraction of primary-particle energy. It was called a long-flying component to distinguish the new component from the penetrating muon component. The search for the long-flying component was performed later by the Pamir experiment, and some excess of cascades has been found at a big depth of lead as well [6].

Two years ago we have reanalyzed data on single hadrons with energies 2 – 40 TeV detected with the use of the BIC and compared their characteristics with results of present-day calculations. The simulation of the absorption of high-energy single (EAS-accompaniment free) protons inside the BIC was done taking for the first time into account the complex calorimeter structure, processes of particle detection with ionization chambers, and cascade selection criteria [7]. The simulation version involving the GEANT 3.21+FLUKA generator describes averaged experimental cascades very well at energies up to 10 TeV. However, the agreement is impaired at energies higher than 10 TeV. A suggestion that a part of single particles incident upon the calorimeter is accompanied by closely-moving low-energy hadrons could describe, more or less, the average hadron cascade curve on the interval 10 – 40 TeV. It was also found in [7] that the EAS-core absorption in lead depends on energy spectrum of EAS-core hadrons very significantly. Therefore the hypothesis that the change in the absorption path of EAS cores can be associated with the change in composition and spectra of EAS-core particles is worthy of study. This is the main goal of this paper.

It is worthwhile to focus attention on the fact that the absorption path starts to rise at EAS-core hadron energies of 40–50 TeV [1]. This corresponds to the primary particle's effective energy of about 2000 TeV [8] that is close to the energy of the so-called "knee" of the primary cosmic-ray (PCR) spectrum. Furthermore the most intriguing features of the PCR spectrum obtained in the recent years are a rigidity dependence of the "knee" energy, i.e., $E_k(Z)=E(1){\times}Z$ ($Z$ is the primary particle's charge) and, perhaps, a very sharp change of the spectrum exponent in this point (by $\delta\gamma \sim$ 2.1) of every nuclear component of PCR [9,10], that is much more than $\delta\gamma \sim 0.5$ in the all-particle spectrum. This result [9,10] is based on the analysis of muon-to-electron ratio in air showers which rises as the EAS energy increases, so the spectrum's steepening measured above the "knee" means the sharp increase of the number of muon-rich air showers that is interpreted as the increase of the portion of showers induced by heavy nuclei. The fact of changing the composition of shower particles is established, but the question that either this is a result of changes in the PCR composition or some other factors cause this change has yet to be answered. In this work we try to answer the following question. Can the abnormal absorption of EAS cores in lead be caused by the change of content of particles in EAS cores and the chemical composition in the PCR? We do not consider here the hypothesis of generation of charmed particles in the interaction of shower particles with lead for the explanation of the result. From one side it requires the very large cross section of charm production and very hard spectra of charmed particles already at energy 10 TeV, from the other side it was described in several papers [1-5].

## 1. Experimental data used for the analysis

The Tien Shan big ionization calorimeter (BIC) of 36-m$^2$ cross-sectional area consists of lead layers interspersed by 5.5×11×300-cm$^3$ copper-wall ionization chambers. The ionization chambers are filled with argon at a pressure of 5 atm. Each array includes an upper lead layer and lower ionization chambers divided by an air gap 5 to 7 cm high [2]. The active readout layers are alternated with the lead layers of thickness 2.5 or 5 cm. The overall thickness of the calorimeter is 850 g cm-2. The total number of the active layers was equal to 15 during the earlier exposure ("old BIC") and 17 during the later one ("new BIC"). For the analysis, we use published data [11] concerning average ionization curves (below the dependence of ionization on calorimeter's depth is called the ionization curve). Besides we use the bank of individual cascades presented by V.I.Yakovlev.

The considered sample was selected by using the following two main criteria: (a) the total measured EAS-core hadron energy over the 36-m$^2$ area, $E_h$, is higher than 1 TeV; (b) the number of EAS electrons is more than 1.3×10$^5$, that, in the average, corresponds to a threshold energy of about

500 TeV. The axes of the air showers are registered within a radius of 3.0 m. Fig.1 presents the total spectrum $dN/dE_h$ multiplied by the factor $E_h^{2.5}$.

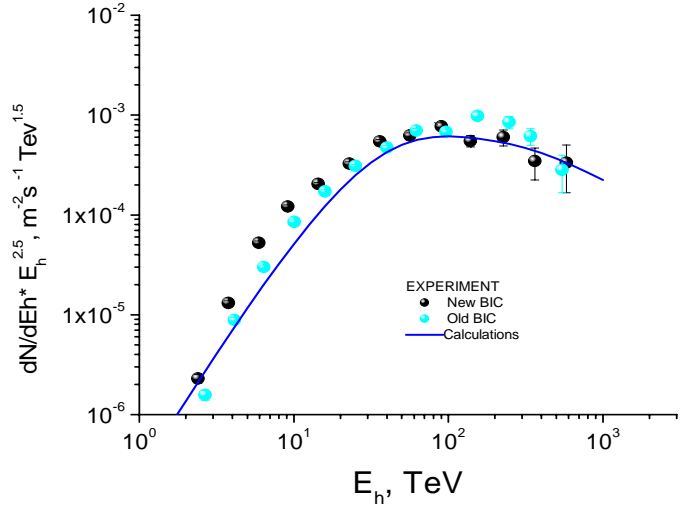

*Fig. 1. Spectra of considered samples of EAS cores selected by the energy $E_h$ from [11] (here $E_h$ is the total EAS-core hadron energy measured over the 36-$m^2$ area): blue circles – "old BIC', black circles – "new BIC'. Line – calculations [8] (CORSIKA+QGSJET02 code with the modern PCR composition). The suppression of the left wing of spectra is caused by the criterion $N_e > 1.3 \times 10^5$.*

In our recent work [8] we have investigated the similar spectrum obtained at the Tien Shan, $F(>E_h)$ but with no using the criterion $N_e > 1.3 \times 10^5$. In that case main features of the $F(>E_h)$ spectrum are a pure power-like form with the index $\gamma = 1.48 \pm 0.03$ at $E_h < 100$ TeV and a noticeable growth of the index $\gamma$ to $1.9 \pm 0.06$ at $E_h > 100$ TeV within a narrow energy interval 1/2 of the order of magnitude wide [8]. In Fig. 1 this "knee" also is evidently observed, but the left wing of the spectrum is suppressed due to criteria $N_e > 1.3 \times 10^5$, which excluded EAS's with energies lower than $600 \pm 200$ TeV. In [8] we used CORSIKA+QGSJET02 code to simulate EAS's at the mountain level.

In Fig. 1 the EAS-core hadron energy spectra calculated also in the framework of CORSIKA+QGSJET02 is marked by line. This version corresponds to the following PCR parameters. Indices of different nucleus group spectra are $\gamma_H = 2.67$, $\gamma_{He} = 2.53$, $\gamma_N$ (other nuclei) $= 2.65$; the rigidity dependent position of the "knee" are $E_k(H) = 3000$ TeV, $E_k(Z) = Z \times 3000$ TeV; the change of indices is $\delta\gamma = 1.2$. As it was shown in [8] the considered set of PCR parameters fits data obtained by the ATIC-2 and *Sokol* direct experiments together with data obtained by the KASCADE experiment for different nuclear groups [12]. The experimental procedure of EAS axes selection and core hadron energy measurement over the 6×6-$m^2$ area was also simulated. This PCR approximation describes the spectrum in Fig. 1 satisfactorily. In accordance with this fit, about 50%, 40%, and 10% of selected EAS cores are originated by protons, helium nuclei, and other nuclei, respectively. All the PCR versions with steep proton and helium spectra ($\gamma > 2.7$ before the "knee") and abundance of heavy nuclei in the "knee" contradict evidently to the experimental spectrum.

Secondly we calculated the effective value of part of energy transferred into the hadron component, $K_{ef}$. At a level of 3340 m a.s.l., $K_{ef} = \langle (E_h/E0)^{1.7} \rangle^{1/1.7}$ at the measured energy $E_h \approx 100$ TeV. If so, $K_{ef}(H) = 0.043 \pm 0.02$; $K_{ef}(He) = 0.024 \pm 0.02$; $K_{ef}(CNO) = 0.017 \pm 0.02$; $K_{ef}(Fe) = 0.0084 \pm 0.02$. Then for the rigidity-dependent position of the "knee" at $E_k(H) = 3000$ TeV we can expect the position of the hadron spectrum "knee": $E_h$ ("knee") $\approx 127$ TeV for protons, $E_h$ ("knee") $\approx 144$ TeV for He nuclei, $E_h$ ("knee") $\approx 306$ TeV for CNO group, $E_h$ ("knee") $\approx 600$ TeV for Fe nuclei. It means that measured hadron energy about 40-50 TeV where the absorption path begins to rise [11] corresponds to the energy of the "knee" if we take into account that only 50–60% of the hadron energy is registered in calorimeter from 4 to 15 layers.

## 2. Simulation of ionization curves in lead

To calculate the EAS core passage through the BIC, we use a version of the FLUKA particle transport code [13]. It must be noted that early versions of the FLUKA hadronic event generator implemented in other codes (GEANT 3, e.g.) have little in common with the present version [14]. The FLUKA is a general-purpose tool for calculations of particle transport and interactions with matter, covering an extended range of applications spanning from proton and electron accelerator shielding to target design, calorimetry, activation, dosimetry, cosmic rays, neutrino physics, radiotherapy etc. The FLUKA hadron-nucleon interaction models are based on resonance production and decay below a few GeV, and the Dual Parton Model (DPM) [14] above. At momenta below 3–5 GeV/c the Peanut package includes a very detailed Generalised Intra-Nuclear Cascade (GINC) and a pre-equilibrium stage, while at high energies the Gribov-Glauber multiple collision mechanism is included in a less refined GINC. Both the modules are followed by equilibrium processes: evaporation, fission, Fermi break-up, gamma deexcitation. The FLUKA can also simulate photonuclear interactions. This code was chosen because some low energy processes (neutron cascade, nuclear disintegrations and so on) playing important role in the formation of ionization in calorimeters are included very accurately.

Because the considered experimental sample spectrum is far from a power-like form, we use monochromatic beams for the analysis: 430 air showers with $E_0 = 10^{16}$ eV (200 p, 100 He, 100 O, 30 Fe) and 500 ones with $E_0 = 10^{15}$ eV (300 p, 200 He). Calculations require a huge computer time, because we have chosen a minimal energy thresholds.

## 3. Comparison of simulated and experimental average ionization curves

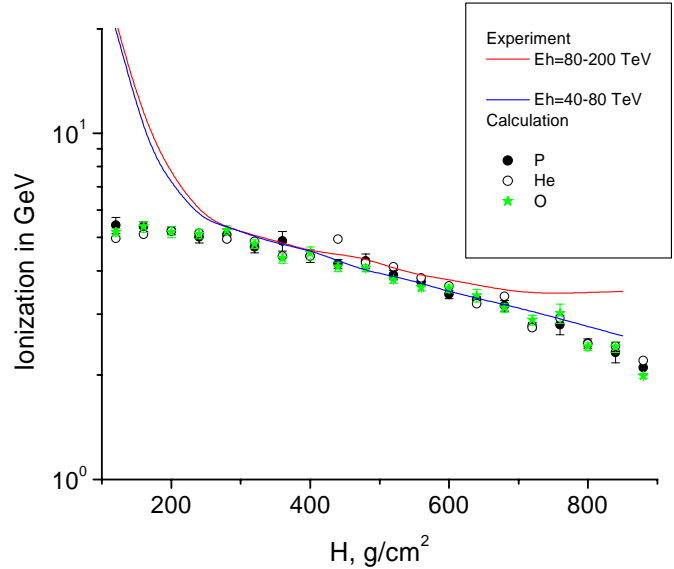

*Fig. 2.* Average ionization curves. Simulated events ($E_h$ =40–200 TeV): black circles (primary protons), open circles (helium nuclei), green stars (oxygen nuclei). Experimental curves: red line ($E_h$ =80–200 TeV), blue line ($E_h$ =40–80 TeV).

Experimental individual EAS cores [11] have been combined into groups by the energy $E_h$ released inside the calorimeter from 4th to 15th layers. Absorption path Λ measured in the interval $H$=340–850 g/cm$^2$ (at $H$ <340 g/cm$^2$ the content of electromagnetic component of EAS cores dominates) increases as $E_h$ rises: $\Lambda=649\pm23$ g/cm$^2$ (at Eh~5,5 TeV), $\Lambda=740\pm27$ g/cm$^2$ (at Eh~9 TeV), $\Lambda=795\pm27$ g/cm$^2$ (at Eh~14 TeV), $\Lambda=912\pm43$ g/cm$^2$ (at Eh~22 TeV), $\Lambda=840\pm33$ g/cm$^2$ (at Eh~34 TeV), $\Lambda=800\pm35$ g/cm$^2$ (at Eh~53 TeV), $\Lambda=865\pm42$ g/cm$^2$ (at Eh~85 TeV), $\Lambda=1000\pm103$ g/cm$^2$ (at Eh~136 TeV), $\Lambda=1300\pm300$ g/cm$^2$ (at Eh~300 TeV). Nearly the same values of $\Lambda$ were obtained with New BIC.

In Fig. 2 we present the similar average curves obtained from the simulated events with energy $E_h$ =40–200 TeV for primary protons, helium and oxygen nuclei. The corresponding calculated absorption paths are following: $\Lambda_p$ =647$\pm$28 g/cm$^2$, $\Lambda_{He}$ = 677$\pm$32 g/cm$^2$, $\Lambda_O$ =626$\pm$16 g/cm$^2$. First, the theoretical absorption path is less than the experimental one in the nearest energy interval; second, we do not observe the noticeable difference in absorption path for various nuclei; third, the large difference in experimental and calculates values of $\Lambda$ is due to last two points corresponding to the large depth of the calorimeter. The difference in theoretical and experimental average curves at $H$<300 g/cm$^2$ is due to contribution of the electron-photon component of EAS, that was not simulated in our calculations.

We investigated as well the average contribution of kaons and neutrons to the ionization curves. One can expect that the part of hadron energy concentrated in kaons attenuates more slowly than the total hadron energy. The kaon-caused absorption path is 860$\pm$40 g/cm$^2$; it seems to be larger than that for the pion component, but the relative contribution of the energy released by kaons does not exceed 10%. The content of muon-produced energy comprises about 2.3% and will be discussed below.

In the next step we start to analyze individual ionization curves.

## 4. Comparison of simulated and experimental individual ionization curves

Further we investigate the dependence of $\Lambda$ measured in individual EAS cores on different characteristics of air showers: a cosine of a zenith angle, a total hadron energy, a hadron energy released between 340 and 900 g/cm$^2$, sort of primary particle, primary energy, model of interactions and so on. It was obtained that the value of absorption path depends negligibly on these parameters. A slight dependence of $\Lambda$ on the hadron energy of EAS cores was observed, but absolute value of simulated absorption path does not exceed 750 g/cm$^2$, that is less than experimental value of $\Lambda$ at high energy.

Besides it was obtained that fluctuations of the value of $\Lambda$ in individual simulated ionization curves become less with growth of core energy, it means that ionization curves become more regular. But many of high energy experimental ionization curves have very irregular shape, as it can be seen from Fig. 3. Individual ionization curves for slow attenuating cores fluctuate very strongly and very often the maximal ionization is released at the end of the ionization curves. Some events look like a 'saw', that reminds the cascades originated by high energy muons [15].

The next parameters (see Fig. 4) were introduced to analyze different types of ionization curves (IC) in relation to the exponential approximation:

1) Depth of the calorimeter where the maximal ionization was produced – Hmax.
2) Value of maximal ionization in relation to the exponential approximation – Ymax.
3) Value of the ionization at the end in relation to the exponential approximation – Yend.
4) Diversity of an ionization in different N layers (Iexp) in relation to the exponential approximation (Iapp): Hi2=$(\sum((\text{Iexp-Iapp})/\text{Iapp})^2/N)^{1/2}$

The last parameter reflects the irregularity of ionization curves. Further we consider these parameters of individual experimental curves in the energy interval $E_h$ =40-200 TeV and individual simulated ionization curves, selected with the same $E_h$ in the dependence on the value of 1/$\Lambda$.

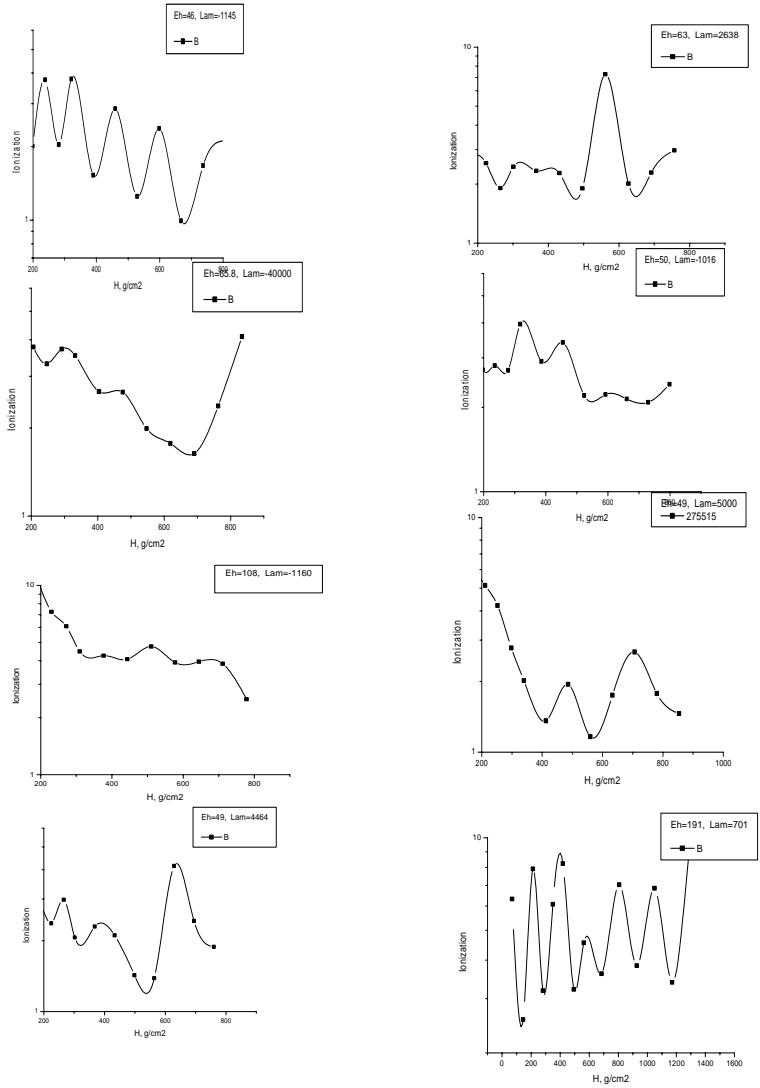

***Fig. 3.*** *The examples of experimental ionization curves with $\Lambda > 1000$ g/cm$^2$ and $E_h > 70$ TeV.*

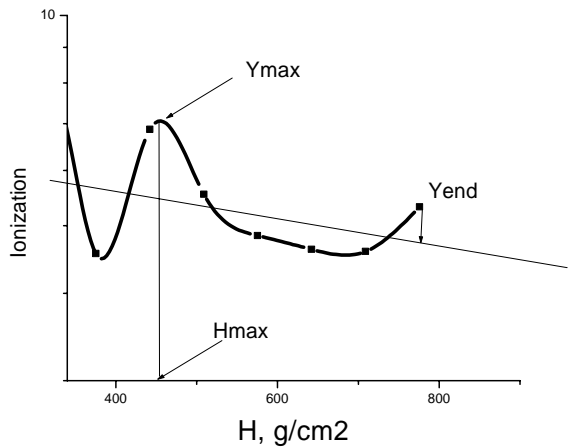

***Fig. 4.*** *Considered characteristics of the individual EAS core ionization curve: line – approximation by exponential function $\exp\{-H/\Lambda\}$ in the interval 340-850 g/cm$^2$; Ymax, Hmax, Yend, Hi2 (see text) .*

In this scale all EAS cores with $1/\Lambda < 0.001$ (including negative values) can be considered as cores with slow energy attenuation in lead. To check the dependence on primary energy also we select two groups with primary energy $10^{15}$ eV and $10^{16}$ eV.

In the Fig. 5 we present the scatter plots Hi2 - $1/\Lambda$, Hmax-$1/\Lambda$, Yend-$1/\Lambda$ for experimental and simulated events. The irregularity of ionizations curves (that is characterized by the parameter Hi2) for the cores with $1/\Lambda<0.001$ is similar to the bulk of cores from primaries with energy $10^{15}$ eV, that is corresponds to small multiplicity of hadrons in the EAS cores. But the value of Hmax and Yend for the cores with $1/\Lambda<0.001$ is noticeably higher than for both groups of simulated cores. They can't be explain by large fluctuations, because in this case we should observe the excess of small value of Hmax and Yend also.

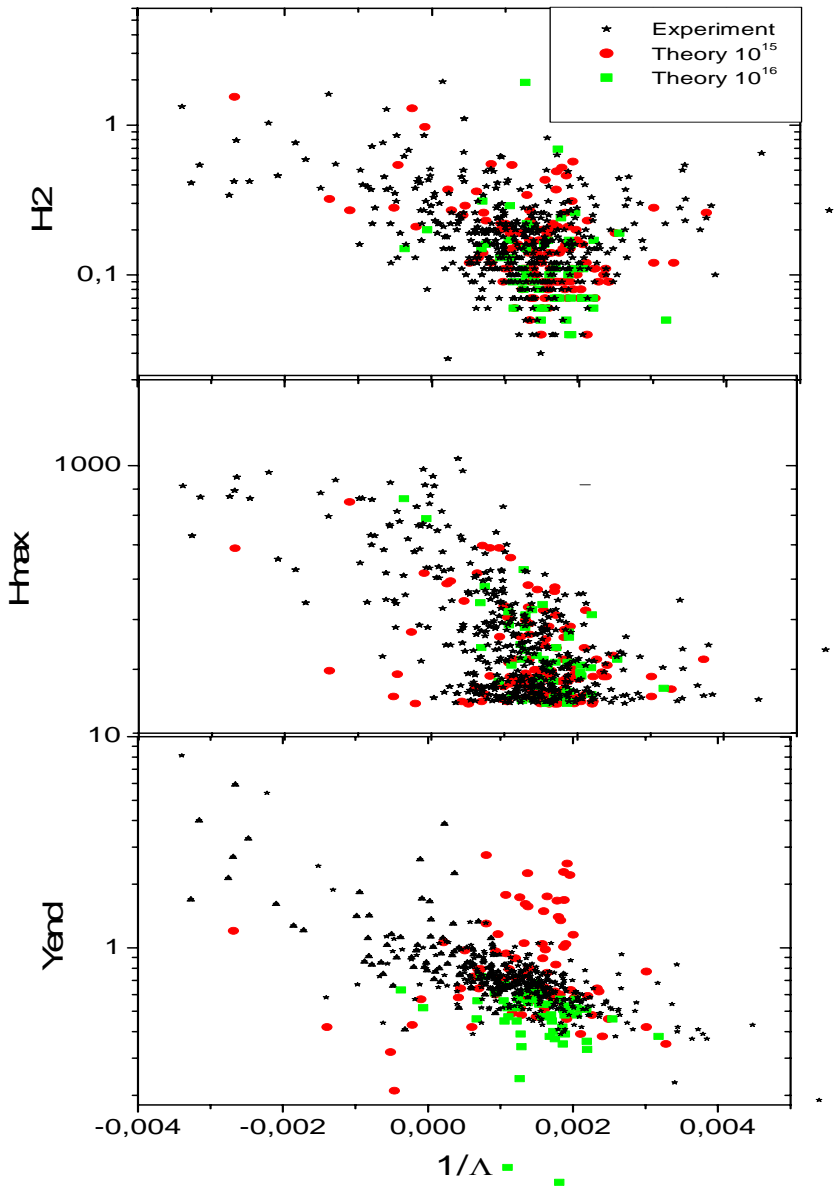

***Fig. 5.*** *The dependence of different parameters Hi2, Hmax, Yend on the inverse value of absorption path $1/\Lambda$ for the experimental cores with hadron energy deposit 40-200 TeV (black stars) and similar samples of simulated cores produced by the primary particles with energy $10^{15}$ eV(read circles) and $10^{16}$ eV (green squares).*

Our calculations show that very strong fluctuations and excess of ionization at large depth of the calorimeter are inherent to ionization curves produced by the group of high energy muons. To demonstrate this conclusion we present in Fig. 6 only muon content of the simulated ionization curves. One can see that groups of muons in air showers explain very well the area of small value of 1/Λ and large value of Hmax, Hi2, Yend. In the next chapter we'll analyze the muon content of EAS in detail.

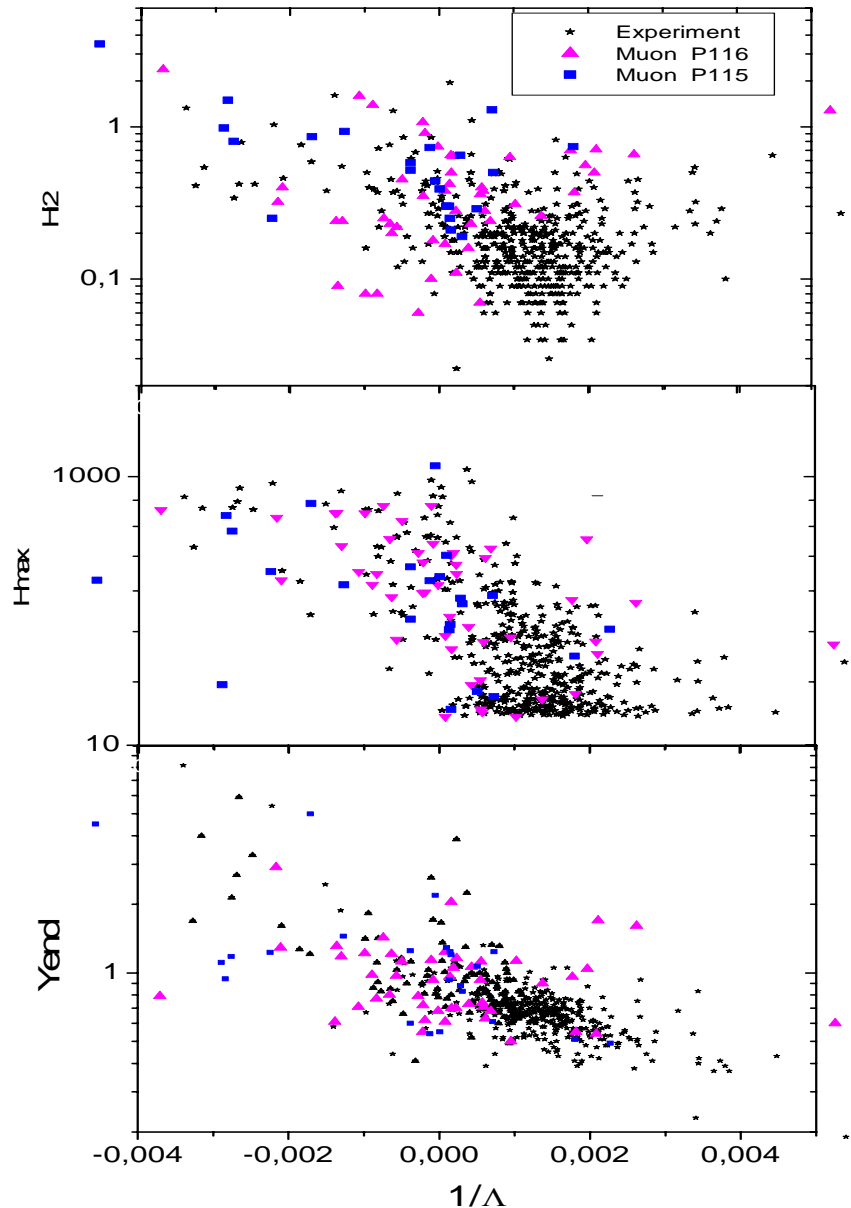

*Fig. 6.* *The dependence of different parameters Hi2, Hmax, Yend on the inverse value of absorption path 1/Λ for the experimental cores with hadron energy 80-200 TeV (black stars) and muon content of simulated cores produced by the primary particles with energy $10^{15}$ eV(pink triangles) and $10^{16}$ eV ( blue squares).*

Comparing the calculated and experimental data we have estimated that the excess of muon-like ionization curves is about 10% at Eh more than 40 TeV.

## 5. Hypotheses for the explanation of the results

In the first instance the general tendencies concerning the content of different particles with energy more than 1 TeV in the EAS cores for different primaries were considered. Our calculation shows that at PeV energy the particles with energy more than 1 TeV carry out the main energy of the core, so these particles determine the fluctuations of ionization curves in calorimeter. 1). The most of EAS core particles are muons and pions. In TeV region kaons can play the important role (up to 30%). 2).The contribution of muon component decreases with the rise of primary energy. 3). The ratio of high energy muons is larger for primary nuclei: for example in the EAS with $E_0=10^{16}$ eV produced by Fe nuclei the number of muons is 20, while in the EAS from proton – only 9 muons. 4). Spectra of hadrons and muons have no a power like form and muon spectra are steeper. 5). While the energy contents of hadron and muon components in EAS cores at the mountain level are compatible, the muon part of energy released in 8-15 layers (just in this interval we measure the value of Λ) in calorimeter already comprises the very small value – several percents, but there are a few cores with the much larger muon contents.

In Fig. 7, we present the distribution on $E_\mu/E_h$ (8-15 layers) where $E_\mu$, $E_h$ is the energy deposit of muons and hadrons between 8 and 15 layers of the calorimeter. The most probable value of $E_\mu / E_h$ (8-15 layers)

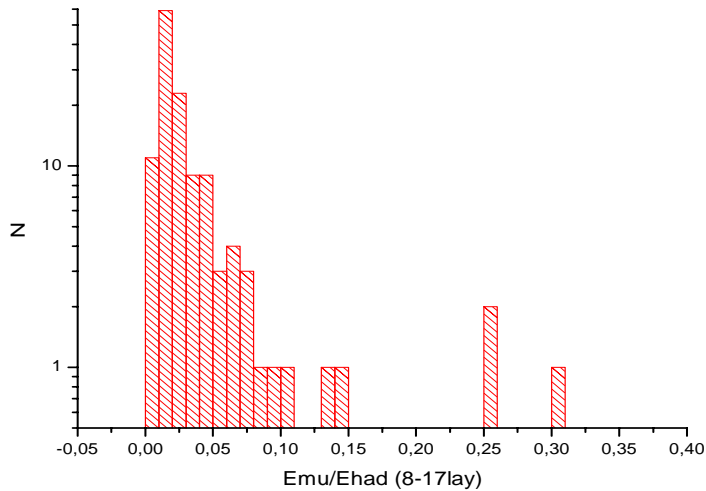

*Fig. 7.* The simulated distribution on $E_\mu /E_h$ (8-15 layers) where $E_\mu$, $E_h$ is the energy deposit of muons and hadrons between 8 and 15 layers of a calorimeter.

The most probable value of $E_\mu / E_h$ (8-15 layers) is about 2.4% - very small, but the distribution on this value has a tail: there is a fraction of EAS cores (~5%) with more than 10% energy deposit produced by muons in the interval 8-15 layers. For these cores absorption path in a half of events exceed 1000 g/cm². This fraction depends on energy (decreases with E), on cos θ (decreases with cos θ), on the distance to the EAS axis (increases with R, if R<5 m), on the sort of primary particles (increases with Z), but couldn't help to explain the 10% excess of muon-like ionization curves even if all primary particles are Fe nuclei.

In the second place we consider the **hypothesis of the so-called prompt muons**. The spectrum of atmospheric muons is more steep than the hadron spectrum due to muon production through pion and kaon decays: with the rise of energy the meson decay length increases, and muons begin to interact but not decay. So the intensity of high energy muons become smaller. Also there is an additional component of muons with slope γ being close to the hadron spectrum [16] ( "prompt" muons produced through charmed and beauty particles decays). The value of energy where the contribution of prompt muons becomes equal to the contribution of esual muons is not known yet and it is intensively discussing now.

Let us consider how muons are detected in our calorimeter. While passing through calorimeter muon creates bremsstrahlung photons and $e^+e^-$ pairs, which give the rise to electromagnetic cascades, detected by ionization chambers. The distribution in relative energies of cascades $u=E_{cas}/E_\mu$ is significantly different for the two processes. At $u>0.1$ the bremsstrahlung process dominates, for $u<0.1$ the probability of pair production exceeds bremsstrahlung by several orders of magnitude. In our work [15] we calculated the spectrum of cascades $I_{cas}(E_{cas})$ initiated in lead by muons with power-like spectra with different exponents: $\gamma=-3.7$ (usual muons), $\gamma=-2.7$ (prompt muons) and for muons with flat spectra $\gamma = -1.7 \div -2$. The coefficient of transition from muon spectrum to the spectrum of cascades with the same energy was calculated: $\eta=I_{cas}/I_\mu$, at $E_\mu=E_{cas}=10$ TeV. This value depends very weakly on energy but very strongly it depends on the slope of spectra $\gamma$. Efficiency at 10 TeV is presented in the Table 1.

**Table 1.** *Efficiency of generation of muon cascades, $\eta=I_{cas}/I_\mu$, at $E_\mu = E_{cas} = 10$ TeV per one g/cm$^2$ for two main processes and for different exponents of power-like spectra of muons.*

| Process \ -γ | 3.7 | 2.7 | 2.0 | 1.7 |
|---|---|---|---|---|
| Brem. Photons | $2.54 \times 10^{-6}$ | $4.18 \times 10^{-6}$ | $7.58 \times 10^{-6}$ | $1.14 \times 10^{-5}$ |
| $e^+e^-$ | $8.09 \times 10^{-8}$ | $6.25 \times 10^{-7}$ | $1.07 \times 10^{-5}$ | $5.59 \times 10^{-5}$ |
| Total | $2.64 \times 10^{-6}$ | $4.80 \times 10^{-6}$ | $1.84 \times 10^{-5}$ | $6.73 \times 10^{-5}$ |

The value of $\eta$ increases with $\gamma$ decrease and besides for very hard muon spectra the pair production process begins to dominate over the bremsstrahlung process. It means that for the more flat spectra of prompt muons the contribution of muon-initiated cascades into the ionization curve will be stressed by the factor of 3-5 additionally to the increase of their number.

We've calculated the absorption of EAS cores in lead supposing that spectra of prompt muons above 1 TeV have the same slope as hadrons, besides we've taken into account the higher efficiency of registration of muons in this case in accordance with Table. 1. In this variant, the simulated $\langle E_\mu/E_h$ (8-15 layers)$\rangle$ value riches value 11% against 2.4% for usual muons, but the excess of slowly attenuating cores with $\Lambda>1000$ g/cm$^2$ couldn't be explained in the frames of the hypothesis of prompt muons.

In the third place we consider another possible source of high energy muons in the atmosphere recently proposed by A.A. Petruhin: that is **"muons from the knee".** [17]. This hypothesis was invented for the explanation of the sharp knee in the primary cosmic ray spectrum. It was assumed [17] that a new process starts to work at the energy around 2-4 PeV and the part of primary energy ("missing" energy) is carried away by a component weakly absorbed in the atmosphere (may be by high energy muons) and not detected by EAS arrays. The shape of the additional muon spectrum can has very flat region around 100 TeV energy [15] due to a sharp start of the new process. Then in the accordance with the Table 1 muons should generate cascades very efficiently and the ratio of muon-like ionization curves should rise. In the hypothesis of A.A.Petrukhin only some minor fraction of EAS with very high energy muons should appear, because they are produced through the new heavy unknown particle which transfer the main energy to muons. But the absent of quantitative parameters of the model does not allow to consider this variant in more detail.

In the forth place we try to analyze this effect from the other point of view. The most complicated for the explanation experimental distribution is that of parameters $H_{max}$ (a depth of a lead, where the maximal ionization was detected) and of the value $Y_{end}$ (relative ionization in 15th layer of the calorimeter) (see Fig. 5). Additional maxima in principal can be produced by low energy processes in lead [7], which was not taken into account. The FLUKA code was specially developed for proton and electron accelerator shielding, calorimetry, activation, dosimetry and other tasks [13] that require very accurate consideration of low energy processes. From the other side some new (or not understand) phenomenon of anomalous neutron bursts detected at Tien Shan station [18] was discovered some years ago. This phenomenon consists in the drastic change (prolongation) of the time distributions of neutron intensity in the events with large neutron multiplicity $M > 1000$. The change of neutron intensity time distributions is accompanied with the corresponding change in the signals of surrounding detectors of the soft (e/g) EAS component. High-multiplicity neutron events are correlated with the passages of the EAS cores with $N_e > 10^6$ [16] corresponded in turn to the energy of the "knee" of primary cosmic ray spectrum. So the threshold energy of two effects (abnormal attenuation of hadrons in lead and neutron bursts) correlates with the energy of the "knee". The study of a possible connection of these two effects will be considered in next paper.

### 6. Conclusion.

In this paper we've performed a reanalysis of the data on EAS core absorption in the lead calorimeter, collected at the Tien Shan mountain station [11]. For this the EAS development in the atmosphere was simulated in the framework of the CORSIKA+QGSJET code whereupon the passage of hadrons and muons through a lead calorimeter has been modeled with using the FLUKA transport code. The calculated dependence of energy deposit of EAS cores on depth of the calorimeter was compared with experimental EAS cores detected at the Tien Shan mountain station with the large 36-m$^2$ lead calorimeter. We analyzed the average absorption of EAS core energy in lead and the characteristics of individual EAS core ionization curves. It was shown that for the EAS with energy around several PeV (that is near the knee in primary CR spectrum) the absorption of energy in lead becomes slower than it is predicted by the modern codes. By this way we confirmed the result obtained many years ago [1]. This effect is connected with the appearance of the excess (about 10%) of abnormal cores with large ionisation released in the lower layers of the calorimeter. It is shown that abnormal EAS cores are most probably produced by high-energy muon groups. A few hypotheses of the excess of muon reach EAS cores are considered. Neither the abundance of heavy nuclei in primary cosmic rays nor the 'prompt' muons originated in the atmosphere can help to explain data.

This work is supported by the RFBR Russian grant 05-02-16781.

**References.**
1. V. I. Yakovlev. Nucl. Phys. B (Proc. Suppl.), v. 122, p. 417 (2003).
2. V.S. Aseikin, et al.. Izvestiya of AN of USSR, ser. fiz., No 5, p. 998 (1974).
3. S.I. Nikolsky, E.L. Feinberg, V.P. Pavluchenko, V.I.Yakovlev. Preprint. FIAN, N 69 (1975).
4. I.M. Dremin, V.I. Yakovlev. Topics on Cosmic Rays. 60-th Anniversary of C.M.G.Lattes. Campinas, Brasil, v.1, p. 122 (1984).
5. I.M. Dremin, D.T.Madigozhin, V.I. Yakovlev. Proc. 21th ICRC, Adelaida v. 10, p. 166(1990).
6. I. V. Rakobolskaya, T.M. Roganova, and L.G. Sveshnikova. Nucl. Phys. B (Proc. Suppl) v.122, p. 353 (2003).
7. L. G. Sveshnikova1, V. I.Yakovlev, A.N.Turandaevskii et al. Physics of Atomic Nuclei, v. 69, No. 2, p. 263 (2006).
8. L.G. Sveshnikova, A.P. Chubenko, N.M. Nesterova et al. Hadron component of EAS cores detected at Tien Shan mountain station in a comparison with CORSIKA + QGSJET simulations. Proc. 14th ISVHECRI, Weihai, China (2006) (submitted to Nucl. Phys. B, Proc. Suppl.)
9. H. Ulrih et al. European Physical Journal C DOI: 10.1140/epjcd/S2004-03-1632-2 (2004)
10. J.R. Hoerandel, Astropart. Phys. V. 19, p.193 (2003).
11. N.M. Nikolskaya, V.P. Pavlyuchenko, V.I. Yakovlev. Parameters of high energy cores of EAS, detected with the large ionization calorimeter on Tien Shan mountain station. Preprint N 42 of Lebedev Physical Institute, Moscow, Russia (1989) (in Russian).

12. M. Antoni T. et al., KASCADE measurements of energy spectra of elemental groups of CR, astro-ph/0505413 (2005).
13. A. Ferrari, P.R. Sala, A.Fass`o, J.Ranft. Fluka: multiparticle transport code. http://www.fluka.org
14. A. Capella, U. Sukhatme, C.-I. Tan and J. Tran Thanh Van. Dual Parton Model, Phys. Rep. v. 236, p. 225 (1994).
15. L.G. Sveshnikova, V.I.Galkin, S.N. Nazarov et al. Phys. Part. Nucl. V. 36, p. 664-666 (2005).
16. L.V.Volkova. Nuovo Cimento C, V.10, N4, p. 465 (1987).
17. A.A. Petrukhin, Proc. 27th ICRC, Hamburg v. 5, p. 1768 (2001); Particle and Nuclei, Letters, v. 6 (109), p. 32 ( 2001).
18. A.P.Chubenko, A.L.Shepetov, V.P.Antonova et al. Proc. 28th ICRC, Tsukuba (2003) v. 69, p.